\documentclass[twocolumn]{aastex61}

\usepackage{appendix}

\accepted{November 28, 2019}
\submitjournal{ApJ}
\shorttitle{Introducing BreakBRDs}
\shortauthors{Tuttle \& Tonnesen}
\begin{document}

\title{BreakBRD Galaxies I: Global Properties of Spiral Galaxies with Central Star Formation in Red Disks } 

\correspondingauthor{Sarah Tuttle}
\email{tuttlese@uw.edu}

\author[0000-0002-7327-565X]{Sarah E. Tuttle}
\affil{University of Washington, Seattle \\
3910 15th Ave NE, Room C319 \\
Seattle, WA, 98195-0002, USA}

\author{Stephanie Tonnesen}
\affiliation{Center for Computational Astrophysics \\
Flatiron Institute \\
New York, NY, USA}

\begin{abstract}

We introduce a collection of primarily centrally star-forming galaxies that are selected by disk color to have truncated disk star formation. We show that common explanations for centrally-concentrated star formation -- low stellar mass, bars, and high-density environments, do not universally apply to this sample.  To gain insight into our sample, we compare these galaxies to a parent sample of strongly star-forming galaxies and to a parent sample of galaxies with low specific star formation rates.  We find that in star formation and color space from ultraviolet to the infrared these galaxies either fall between the two samples or agree more closely with galaxies with high-specific star formation rates. Their morphological characteristics also lie between high- and low-specific star formation rate galaxies, although their Petrosian radii agree well with that of the low-specific star formation rate parent sample.  We discuss whether this sample is likely to be quenching or showing an unusual star-formation distribution while continuing to grow through star formation.  Future detailed studies of these galaxies will give us insights into how the local conditions within a galaxy balance environmental influence to govern the distribution of star formation. In this first paper in a series, we describe the global properties that identify this sample as separate from more average spiral galaxies, and identify paths forward to explore the underlying causes of their differences.

\end{abstract}


\section{Introduction} \label{sec:intro}
Over the last two decades, galaxy population studies at low to intermediate redshift have exposed a fundamental evolution in the overall galaxy population since a redshift of z$\sim$2 \citep{2004ApJ...608..752B, 2007ApJ...665..265F,2013ApJ...777...18M, 2013A&A...556A..55I, 2014ApJ...783...85T,  2016A&A...590A.103M}.  However, isolating the processes responsible for this shift in the galaxy population has been a great deal more complex than identifying the evolution.

To go from a blue star-forming galaxy to a red quiescent galaxy requires the cessation or truncation of star formation. Understanding how this cessation occurs and how its timing percolates through a galaxy is crucial to identifying the mechanisms that drive galaxy evolution. Galaxy quenching can take on many forms, fundamentally segregated by galaxy mass, morphology, and environment \citep{2010ApJ...721..193P}.  The influence of galaxy morphology may be studied by observing global properties of the stellar populations of galaxies. Most galaxies can be described by a spheroid component consisting mainly of an old stellar population, and may also feature an extended disk with continued star formation that contains younger stars (as discussed in, e.g., \cite{2013MNRAS.430.2622D}).  This two-component description of galaxies has been applied in different incarnations for more than 100 years (e.g. \cite{1926ApJ....64..321H}). 

An observed age difference between bulges and disks implies an ``inside-out" formation mode, in which the bulges of spiral galaxies form early with little to no current star formation, while star formation continues until later times in their disks  \citep{1991ApJ...379...52W,1997ApJ...477..765C, 1998ApJ...507..601V, 1999MNRAS.307..857B}. In spiral galaxies, inside-out behavior has been observed across evolutionary time and with many techniques, frequently tied to a transitional mass \citep{2005ApJ...630L..17T, 2011MNRAS.412.1081W, 2013ApJ...778L...2C, 2015ApJ...804L..42P} Recently, integral field spectroscopy has been used to better explore star formation histories and how they differ across individual galaxies. Observations with MUSE and CALIFA have explored specific star formation through the use of resolved gas and stellar populations, respectively \citep{ 2013ApJ...764L...1P,2018A&A...615A..27L, 2019MNRAS.484.5009E}. They find results consistent with the inside-out path for spiral galaxy growth. In work with p-MaNGA, star-forming galaxies do not show a strong radial dependence for star formation, but centrally quiescent galaxies have a negative gradient in support of the inside-out premise \citep{2015ApJ...804..125L}. 

While many spiral galaxies fit this picture of an old bulge within a star-forming disk, there is evidence that inside-out growth is not universal, particularly at low masses %. Mass may also play a role in low-mass galaxies that have been found to grow from the outside-in  
\citep{2008ApJ...682L..89G}.  Indeed, \citet{2013ApJ...764L...1P} found that outside-in growth can occur up to M$_*\sim$ 10$^{10}$ M$_{\odot}$.  \citet{2015ApJ...804L..42P} find similar trends with regards to stellar mass in color gradients of galaxies, while \citet{2016MNRAS.463.2799I} find a large diversity in radial age gradients for lower mass systems. Simulations that focus on low masses (M$_*\leq$ 10$^{10}$ M$_{\odot}$) show star formation may preferentially occur in the central regions of a galaxy, with feedback from supernovae shaping the overall mass profile (e.g. \citet{2010Natur.463..203G}; \citet{2016ApJ...820..131E}). When we compare observational results to simulation outcomes, it becomes clear that we have not isolated the key processes that are shaping spiral galaxy evolution \citep{2018ApJ...866L..21P,2018arXiv181201017S, 2019MNRAS.484.4413H, 2019MNRAS.tmp..741T}. 

This mass segregation and mismatch between simulation and observation may occur because different physical processes appear dominant at different scales. Massive galaxies are shaped by galaxy-wide processes - active galactic nuclei disrupting star formation on large scales, and mergers obliterating the morphologies and structures that shape spiral galaxies (e.g. \citet{2015ARA&A..53...51S} and references therein). Smaller star-forming galaxies tend to be influenced more strongly by their environment through processes such as tidal gas stripping, ram pressure stripping, starvation, or strangulation \citep{2006PASP..118..517B}. The influence of each of these processes rises and falls in any galaxy depending on many variables including its clustercentric (groupcentric) radius, total galaxy mass, stellar mass, and mass surface density (e.g. \citet{2010ApJ...721..193P, 2004MNRAS.353..713K, 2018MNRAS.477.3014B, 2019MNRAS.tmp..741T}. Secular processes caused by galaxy characteristics like the existence of a bar \citep{2004ARA&A..42..603K,2012ApJS..198....2k, 2013ApJ...779..162C}, winds, or disk heating \citep{2003AJ....126.2707S,2016ApJ...826...59W} may have a strong influence in instances where outside processes haven't yet interfered \citep{2014MNRAS.438.1552F}.

While most galaxies to z$\sim$1 are either blue, star-forming late-type galaxies or red early-type galaxies \citep{2006MNRAS.373.1389C, 2009A&A...493...39M}, a significant number of galaxy ``classes" have arisen that violate (or perhaps expand) our understanding of the connection between galaxy color (stellar content) and its morphology (shaped by galaxy dynamics and external interactions). Galaxies have been identified dwelling in between other classifications, even as early as the ``anemic spirals" of \citet{1976ApJ...206..883V}. These galaxies were found to be spiral-like, but much less vigorously star-forming than their traditional counterparts due to their dense environments, and were suggested as a transitional classification. This connection tying galaxy environment directly to the transition of individual galaxies has been demonstrated repeatedly (e.g. \citet{1999ApJ...518..576P, 2004ApJ...601..197P}) including the identification of blue passive galaxies in rich clusters (indicating recently truncated star formation \citep{2009MNRAS.400..687M}). Some galaxy classes, like the passive red spirals discovered via Galaxy Zoo, may show a less clear connection with their environment. \cite{2010MNRAS.405..783M} find that this population (making up 30\% of spiral galaxies) is not correlated with environment, while \citet{2009MNRAS.393.1324B} found that the fraction of passive spirals was dependent on environment.
This decoupling of morphological transformation from the quenching of star formation was recently shown in local galaxies by \citet{2016MNRAS.462L..11F}.

Another class of galaxies that does not fit into classical categorization are galaxies morphologically identified as early-type may have star formation. \citet{2010ApJS..187..374S} examined optical color profiles of nearby early-type galaxies and found that $\sim$30\% of them show positive color gradients, evidence for central star formation. \citet{2015A&A...583A.103G} find that star-forming blue early-type galaxies may indicate that once-quenched galaxies can be rejuvenated by tidal interactions. Indeed, some galaxies have been observed to fall outside of standard morphological categories in a way that may indicate a transitional nature, such as the recent ``red misfits" of \citet{2018MNRAS.476.5284E}. These misfits tend to be massive star-forming galaxies with active galactic nuclei (AGN). 

One promising approach to determine how galaxies move from star-forming to quenched is to focus not just on outliers but on potential transitional galaxies.  Green valley galaxies are defined as galaxies that reside between the blue cloud and red sequence in stellar mass versus color space \citep{2007ApJS..173..342M}. Originally thought to be star-forming galaxies which were fading, their structural differences imply other likely mechanisms such as building up of the bulge \citep{2014SerAJ.189....1S}.  Work from \citet{2017ApJ...851...18L} suggests that the fading of green valley galaxies is driven by a dropping gas fraction, which is consistent with earlier results \citep{2014MNRAS.440..889S, 2015MNRAS.450..435S}. Observations show green valley galaxies are more centrally concentrated than other galaxies of the same mass \citep{2007ApJS..173..315S}. Green valley galaxies that show suppressed star formation also appear to be quenched globally, not solely within the disk \citep{2018MNRAS.477.3014B,2018MNRAS.474.2039E}.  There have also been some observations of green valley galaxies, particularly massive ones, that indicate that the sSFR may be more depressed in the center than in the outskirts \citep{2016A&A...590A..44G, 2016ApJ...828...27N, 2018MNRAS.475.5194M,2018MNRAS.477.3014B,2018arXiv181208561M}.  Whether the presence of a massive bulge has a strong impact on the SFR of galaxies remains an open question--\citet{2018MNRAS.475.5194M} find that possession of a large bulge does not consistently impact SFRs, consistent with results from \citet{2019ApJ...874..142K}, while \citet{2016A&A...590A..44G} and \citet{2017MNRAS.466.1192M} find that galaxies with higher bulge fractions have lower SFR for their stellar mass surface density.

What tools can we use to try to disentangle these influences on galaxy evolution? Using the minority of galaxies that fall outside standard color-morphology relationships is one approach suggested by the data. For example, red late-type galaxies in clusters indicate that color evolution occurs on a shorter timescale than morphological evolution \citep{1999ApJ...518..576P, 2009MNRAS.393.1324B}. Using Cosmic Evolution Survey data, \citet{2010ApJ...719.1969B} argue that as much as 60\% of spiral galaxies move through a passive spiral phase on the way to the red sequence. These deviations provide opportunities for insight.

In this paper, we identify a class of galaxies that falls outside of the red-bulge, blue-disk (or inside-out) formation model:  primarily centrally star-forming galaxies that are selected by disk optical color to have truncated disk star formation. %The initial hypothesis was that dense environments would lead to truncated star formation in disks while leaving central star formation ongoing. Instead we found galaxies which met the selection criteria, but are present in all environments. Either we are selecting galaxies which experience different processes depending on their environments that truncate outer star formation, or the main cause for the cessation of the star formation in these disks is not environmental.
This small sample of star-forming galaxies is well-fit by bulge/disk decomposition \citep{2012MNRAS.421.2277L}(from here forward LG12) and presents with star-forming centers and red ($g - r$) disks. 

We begin by introducing the sample in Section \ref{sec:sample}. We call these galaxies ``breakBRDs" (break Bulges in Red Disks) because they are selected for central star formation via their D$_n$4000 break values, hence ``break Bulge", and have optically red disks. We characterize the parent sample, and remove AGN using cuts on the BPT diagram. In Section \ref{sec:comps} we further divide our parent sample into galaxies with high- and low-specific star formation rates (sSFR; SFR/M$_*$) so we can better use them to determine whether our sample is star-forming or quenching. Section \ref{sec:MME} compares our sample to the high- and low-sSFR parent samples in the mass, environment, and morphological parameter spaces.  In order to understand their star formation history in more detail, in Section \ref{sec:SFH} we examine ultraviolet, optical, and infrared colors as well as stellar mass.  % to attempt to isolate them in evolutionary space. We move on to examine the star formation histories in Section \ref{sec:SFH}. %In this paper we introduce a small sample of star-forming galaxies well-fit by bulge/disk decomposition \citep{2012MNRAS.421.2277L}(from here forward LG12) that present with star-forming centers and red ($g - r$) disks. We begin by introducing the sample in Section \ref{sec:sample}. We call these galaxies ``breakBRDs" (break Bulges in Red Disks) because they are selected for central star formation via their D$_n$4000 break values, hence ``break Bulge", and have optically red disks. We characterize the parent sample, and remove AGN using cuts on the BPT diagram. In Section \ref{sec:comps} we further divide our parent sample into galaxies with high- and low-specific star formation rates (sSFR; SFR/M$_*$) so we can better use them to determine whether our sample is star-forming or quenching. Section \ref{sec:MME} compares our sample to the high- and low-sSFR parent samples in the mass, environment, and morphological parameter spaces.  In order to understand their star formation history in more detail, in Section \ref{sec:SFH} we examine ultraviolet, optical, and infrared colors as well as stellar mass.  % to attempt to isolate them in evolutionary space. We move on to examine the star formation histories in Section \ref{sec:SFH}. 
We also consider their HI reserves in Section \ref{sec:HI}.  Moving briefly from global properties, in Section \ref{sec:fiber} we focus on the spectral measures from the central SDSS fiber.  In Section \ref{sec:disc}, we discuss possible scenarios that may explain the
 data underlying this sample.  We summarize our findings and discuss future work in this series in Section \ref{sec:conclusion}.

\section{Sample Selection} \label{sec:sample}
The sample studied here is selected from a large local ($z < 0.05$) sample of face-on SDSS galaxies used for bulge/disk decomposition by LG12. The source catalog was specified to be face-on, which removed the potential for confusion of star-forming regions located in the disk overlapping with the center of the galaxy. Their morphological appearance is diverse, as is discussed further in Section \ref{sec:MME}. 

Our sample is derived from LG12, in which the authors develop an astrophysically-guided bulge/disk decomposition approach. They then select galaxies with r-band apparent magnitudes brighter than 17.7, and require galaxies to be face-on galaxies by limiting the axial ratio to $ > 0.25$ as measured by the SDSS pipeline. This improves the likelihood of success for bulge/disk decomposition and avoids dust lane contamination. Their 71825 galaxies are drawn using a low-redshift sample ($0.003 < z < 0.05$) from the NYU Value-Added Catalog (VAGC)\citep{2005AJ....129.2562B, 2009ApJS..182..543A}. LG12 perform their decomposition on DR8 images (NYU VAGC was created with DR7) due to a significant improvement in sky subtraction between SDSS DR7 and DR8.

\subsection{Parent Sample}

For this project, we downselected from the LG12 sample using several cuts to search for galaxies with the most robust bulge/disk decompositions.  The following criteria were therefore applied: requiring the r-band absolute magnitude as measured in the fit to be $M_r < -19$, limiting the range of $g - r$ colors in both the bulge and disk to $0.2 < g - r < 0.9$, and limiting the axial ratios to $> 0.7$ (Lackner, personal communication). Using this criteria, the parent sample contains 4643 galaxies, the majority of which contain red bulges with a red or blue disk (as one might expect).

\subsection{Selecting Red Disks}\label{sec:reddisks}

From the parent sample we searched for galaxies with central star formation and red disks. Red disks were determined as those with $g - r > 0.655$ using either a de Vaucouleurs bulge and an exponential disk (nb4 model), or an exponential bulge and exponential disk (nb1 model).  We used these two fits on all galaxies to find our sample, regardless of their ``best fit'' model from LG12 (we have the ``best fit" model for 121 of our 126 galaxy sample from Lackner, personal communication).  In this section, we discuss the LG12 fitting procedure below, and how many of our breakBRD galaxies fall into each ``best fit'' model. For more details, we refer the reader to LG12.

In LG12, five fits were used for every galaxy:  a pure exponential disk, a single de Vaucouleurs profile, the two bulge + disk models (B+D), or a single component Sersic model.  Selecting which of these models is the best fit for a galaxy is non-trivial, as explored in detail in LG12.  This is largely because $\chi^2$ values generally do not differ greatly between models, and due to structure in the galaxy (bar, rings, spiral arms, etc.) the $\chi^2$ values are often quite high.  Therefore, much of their selection is based on astrophysically-motivated choices to separate pseudo-bulges and classical bulges, as well as galaxies where a bulge + disk fit was poorly suited (often due to a galaxy being blue and faint).  However, we are searching for a galaxy sample that upends our assumptions about galaxy growth and evolution, so here we carefully consider each criteria and whether a bulge + disk fit is a truly unphysical model.  

LG12 chose a single component Sersic model as the ``best-fit" model unless the galaxy was better fit by one of the other four models, either using the  $\chi^2$ values or via astrophysical selection.  Therefore, in LG12 the Sersic ``best-fit" category includes the most galaxies, especially the (intrinsically) faintest galaxies in the sample, irregular galaxies, strongly barred galaxies and galaxies with otherwise poor model fits.  We highlight the fact that galaxies with bulges much bluer than their disks are likely to have strong bars, which could drive the central star formation for which we are searching.  Therefore we must allow galaxies with a Sersic ``best fit'' model into our sample.  In fact, 21 of the 126 galaxies in our sample are best fit with a single component Sersic model. 

To find most of their exponential disks, LG12 select galaxies for which the disk in the B+D model matches the single-exponential fit in total flux, axial ratio of the fit ($q_d$) and R$_{eff}$ to within 10 per cent.  As expected, no galaxy that is best fit by a pure disk model (with no bulge component) falls into our sample.

In order to find galaxies best fit by a single de Vaucouleurs model, likely elliptical galaxies, LG12 study the colors and morphologies of the few galaxies with low $\chi^2$ values for the de Vaucouleurs fit.  Based on this small set of galaxies, LG12 use total $g -r >$ 0.55, $g -i >$ 0.80, and $b/a > 0.55$ to find more galaxies along the red sequence that may be best fit by the de Vaucouleurs model.  In order to only include likely elliptical galaxies and not S0s, they then require that the exponential bulge and disk (nb1) fit finds a round ($q_{disk} > 0.4$), red ($g -r >$ 0.65) disk, and the deVaucouleurs bulge and exponential disk (nb4) fit finds a large (B/T $>$ 0.4), red ($g -r >$ 0.65), round ($q_{bulge} >$ 0.55) bulge.  These criteria clearly could include face-on galaxies with red disks and star-forming bulges, so we must not eliminate galaxies with this ``best fit'' model.  13 of our galaxies are best fit by this model.

In order for a galaxy to be best fit by a bulge + disk model, the bulge and disk in both models must be detected in g, r, and i-band images.  LG12 also require that the bulge R$_{eff}$ be smaller than that of the disk, and that the bulge flux dominates in the central part of the galaxy.  Finally, the bulge and disk must have similar axial ratios (within a factor of two).  The authors use purely astrophysical arguments to distinguish between de Vaucouleurs fit classical bulges (nb4) and exponential fit pseudobulges (nb1).  If the nb4 bulge component has $g -r >$ 0.6 and $q_{bulge}/q_{disk} >$ 0.65 the classical bulge fit(nb4) is used, otherwise the exponential bulge fit is chosen to be the ``best fit''.  Most of the breakBRDs are best fit by one of these bulge + disk models, 87 out of 126.  Because we are selecting galaxies with central star formation, most of our bulges are blue and therefore the model chosen by LG12 is that with an exponential bulge (79).  

Galaxies were selected to be breakBRDs as long as they fulfilled all the selection criteria using either bulge + disk fit. 
In total, 92 galaxies are chosen using the de Vaucouleurs bulge fit and 78 galaxies using the exponential bulge fit, with an overlap of 44 galaxies chosen with both fits.  As we show in Figure \ref{fig:parentsamplereddisk}, using the de Vaucouleurs bulge and an exponential disk (nb4 fit) most of the galaxies in our sample present red disks.  We reiterate that whatever the ``best fit" chosen in LG12, all galaxies in LG12 were fit using every available model, and in this work breakBRD galaxies are selected using a bulge + disk model.

Out of our 4643 parent galaxy sample, 3820 galaxies have red disks using either of the bulge + disk fits. 
\subsection{Searching for Central Star Formation}

From the sample of galaxies with red disks, we select galaxies with central star formation. Central star formation is found using the D$_n$4000 diagnostic in the central fiber of the SDSS spectrograph as discussed in \citet{2013MNRAS.428.2141L}(LG13). We use the D$_n$4000 from \citet{2004MNRAS.351.1151B}(B04), which uses the narrow definition of D$_n$4000, from \citet{1999ApJ...527...54B}. LG13 find that two-thirds of their galaxies have a bulge-to-total flux ratio within 3 arcsec (the size of SDSS fibers) larger than 0.5. Therefore they find that the fiber quantities are typically dominated by the stellar light from the bulge.  

\begin{figure}
%\epsscale{1}
\includegraphics[scale=0.53]{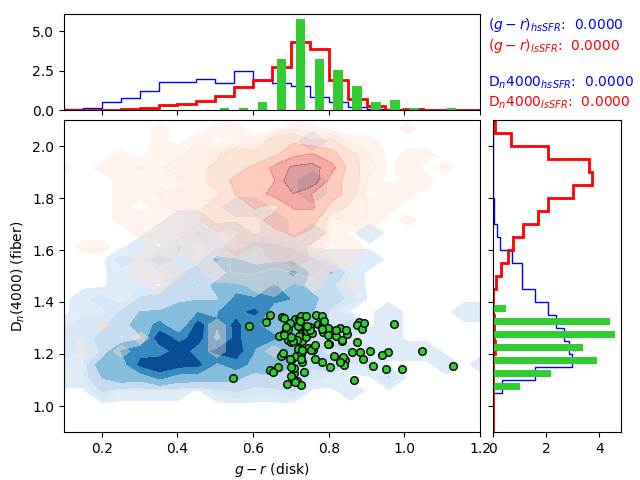}
\caption{We plot our selection criteria, showing the (g-r) disk color (using a de Vaucouleurs fit for the bulge and an exponential disk, or nb4 fit) compared to the $D{_n}4000$ value from the SDSS fiber spectrum. BreakBRD galaxies are the green circles and histograms. The parent sample is shown with the underlying contours and unfilled histograms, with low- and high-sSFR samples shown in red and blue, respectively. The low-sSFR histograms are a heavier line to aid the eye.  The p-value from the two sample KS test between the BBRD galaxies and the low- and high-sSFR samples are listed in the upper right corner for the plotted variables.  The p-value will be reported for all histograms in their respective figures.
Even when holding the fit constant (nb4), the sample selected contains galaxies with much redder disks than comparably star-forming galaxies in the parent sample. \label{fig:parentsamplereddisk}}
\end{figure}

In their Figure A3, LG13 show that there are two populations of bulges, those with star formation within the last $\sim$1 Gyr and those without recent star formation, with small and large D$_n$4000, respectively. LG13 fit the distribution of D$_n$4000 with two Gaussians and assign each galaxy a probability of having a classical quiescent bulge or star-forming pseudobulge based on the ratio of the Gaussians at a given D$_n$4000. If the bulge is more likely to be star-forming we include it in our sample. As we see in Figure \ref{fig:parentsamplereddisk}, the D$_n$4000 values of the breakBRD galaxies are all below 1.4, indicating recent star formation in simple models \citep{2003MNRAS.346.1055K}.  

We find galaxies with star-forming bulges in red disks to be 2.7\% of the parent sample.

\subsection{Removing AGN}\label{sec:AGN}
\begin{figure}
\includegraphics[scale=0.6]{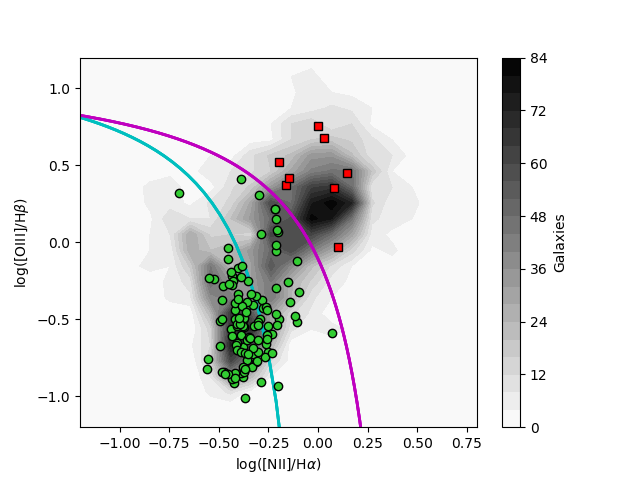}
\caption{Here we use the BPT diagram, used to identify the dominant ionization mechanism in the nebular emission lines of galaxies, to separate out AGN from composite and star-forming galaxies. The underlying black contours show the distribution of the parent sample. The magenta line and the turquoise line are adopted from \citet{2002ApJS..142...35K} and \citet{2003MNRAS.346.1055K} respectively to distinguish between AGN activity above the magenta line, star-forming activity below the turquoise line, and composite activity between the two. The overplotted points show our galaxy sample to be primarily star-forming, with all selected breakBRDs indicated by green circles. Eight galaxies are selected as AGN (red squares).}\label{fig:bpt}
%\end{center}
\end{figure}

One expected population present in this sample are AGN. Although they have blue bulges, they do not give us insight into our question about ongoing star formation in the bulge versus a quiescent disk, so we remove them from our sample. The BPT diagram is used to separate out likely AGN via the methods described in \citet{2002ApJS..142...35K}, \citet{2003MNRAS.346.1055K}, and \citet{1981PASP...93....5B}. Figure \ref{fig:bpt} shows the result of this classification, using the emission line strengths from B04. The points show our sample while the contours demonstrate the distribution of the parent sample. The galaxies in Figure \ref{fig:bpt} fall into three broad regions-- star-forming below the first aquamarine line, composite between the two lines, and AGN above the magenta line. We conclude from the BPT diagram that only 8 of the 126 selected galaxies are AGN. These are shown with red squares in Figure \ref{fig:bpt} and are excluded from further analysis. Our remaining galaxies lie in either the star-forming or composite regime (green circles). This final sample of 118 galaxies is the breakBRD sample.

For consistency, we also select only star-forming and composite galaxies in the parent sample for the comparisons we discuss below. This leaves 2499 galaxies, and we note that more than 2000 of the 2144 galaxies that have AGN come from the red disk sample identified above (Section \ref{sec:reddisks}).

\subsection{Comparing to Strongly and Weakly Star-Forming Galaxies}\label{sec:comps}

We have selected a unique galaxy sample using the radial distribution of star formation, but must consider other properties to determine what correlates with or possibly drives the inner star formation in red disks. As this sample falls outside of the canonical picture of galaxy growth and evolution, we split the comparison sample into high- and low-sSFR galaxies, using the minimum of the sSFR histogram at sSFR $ \sim10^{10.9} M_{\odot} yr^{-1}$ (see Figure \ref{fig:massSFRs}) (1280 and 1219 high- and low-sSFR galaxies, respectively). We are then able to compare our sample separately to strongly star-forming galaxies and those that are transitioning or quenched. This may indicate whether our sample is of regularly star-forming galaxies with unusually centralized star formation or of galaxies that are quenching with their final star formation in the central regions.

As illustrated in Figure \ref{fig:parentsamplereddisk}, the D$_n$4000 values of our sample align well with strongly star-forming galaxies in the parent sample, but the disk colors are generally much redder, and agree better with the disk colors of galaxies with low-sSFRs. However, we performed two-sample Kolmogorov-Smirnov (KS) tests on each of these parameters and find that the disk $g - r$ colors and D$_n$4000 values of our sample are different from both the low- and high-sSFR parent samples.

\section{Mass, Environment, and Morphology}\label{sec:MME}

Morphological transformation through quenching may be driven by mass and environment \citep{1980ApJ...236..351D}. For example, central star formation in otherwise red disks is observed in ram pressure stripped galaxies in clusters \citep{2004ApJ...613..866K,2006PASP..118..517B,2017Natur.548..304P}. We first try to identify the influences on our sample galaxies by examining their mass, morphology, and environment.  

\subsection{Mass}

%fig 
\begin{figure}
%\begin{center}
\includegraphics[scale=0.53]{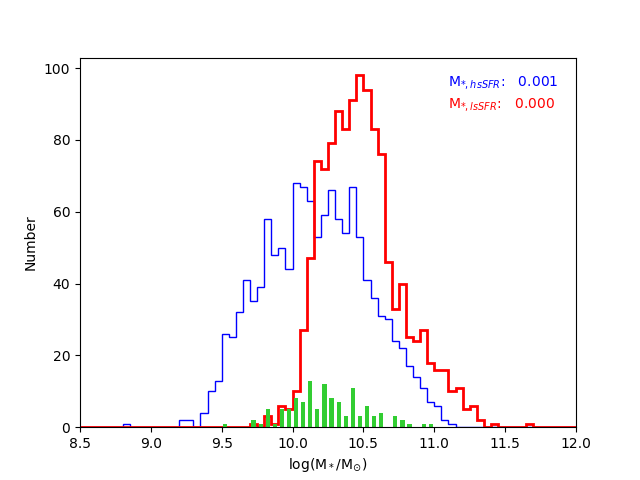}
\caption{The mass histograms of the breakBRD sample in green compared to the high and low-sSFR parent samples in blue and red, respectively.  The p-values from comparing the two parent samples to the breakBRD sample are shown in the figure.}
\label{fig:massden}
\end{figure}

Current studies indicate quenching is strongly shaped by mass, with high mass galaxies more likely to be quenched (e.g. \cite{2010ApJ...721..193P}).  We also see this correlation by comparing the high- and low-sSFR parent samples in Figure \ref{fig:massden}, shown in blue thin and red thick lines, respectively.  However, the breakBRD sample does not easily fall into a high-mass quenched sample or a low-mass star-forming sample, and instead lies between the mass distributions of the high- and low-sSFR parent samples (as quantitatively shown using KS test p-values).

Neither does the mass distribution of breakBRD galaxies agree with centrally-concentrated star-forming galaxies.  As discussed in the Introduction, centrally-concentrated star formation is thought to occur more often in low-mass galaxies (e.g. \citet{2013ApJ...764L...1P}; \citet{2010Natur.463..203G}).  More than half of the galaxies in the breakBRD sample have stellar masses above 10$^{10} M_{\odot}$, well above the stellar masses at which this phenomenon has been observed.

Because many properties of galaxies correlate strongly with stellar mass, differences between breakBRD galaxies and the parent samples could be traced to these dramatically different mass distributions.  Therefore, in the rest of this paper we mass-weight the parent samples so as to eliminate this bias without losing any information from the larger samples (as may be the case using a sub-sampling method). We use the mass histograms and assign every parent galaxy a weight defined as the ratio of the number of breakBRD galaxies to the number of parent galaxies in a mass bin.  We do this separately for the high-sSFR and low-sSFR parent samples.  This results in mass distributions whose weighted KS test has a p-value of $\geq$ 0.99.  We then use the weighting from the mass distributions for all the other features we plot using the parent samples.  We note that we perform this mass weighting on every subsample of breakBRD galaxies: the entire sample from the SDSS galaxies, the smaller subsample with WISE data, and the BPT-selected star-forming galaxies.

\begin{figure}
\includegraphics[scale=0.53]{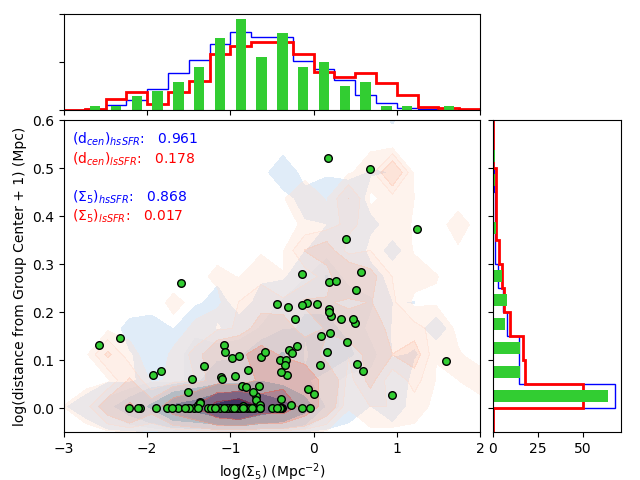}
\caption{$\Sigma_5$ versus distance from the group center, comparing two different measures of environmental density.  Our sample does not deviate from the underlying parent distribution and the galaxies are distributed throughout environments from being central group members to galaxies on the outskirts. The underlying sample is separated by specific star formation rate, as in Figure \ref{fig:parentsamplereddisk}. \label{fig:rho5cd}}
\end{figure}

\subsection{Environment}
Examining galaxy environment more closely in Figure \ref{fig:rho5cd} (using the measures from LG13), we clearly see that environmentally-driven evolution should not preferentially effect our sample of galaxies, as both their local density and satellite fraction is similar to the parent sample.  The y-axis shows the log of the distance from the group center plus one Mpc, which is 0 when a galaxy is the central galaxy in a group.  The breakBRD galaxies are distributed throughout all environments as measured using the local galaxy density, $\Sigma_5$, using the nearest five neighbors on the sky. 

Quantitatively, we find that 40\% of the galaxies in our sample are centrals, versus 39\% of the comparison sample (46\% of the high-sSFR galaxies and 32\% of the low-sSFR galaxies), and satellite galaxies are distributed at a range of distances from the group center, out to a radius of more than 2 Mpc. Indeed, the KS tests indicate that the environments of the breakBRD sample may be from the same distribution as either the high- or low-sSFR parent samples (we note that the p-values are quite small when comparing the high- and low-sSFR samples for either variable).  Thus we find that breakBRD galaxies appear in high- and low-density environments, near or far from the group center with no differentiation from the parent sample. 

Because environment is correlated with many galaxy properties, we have performed many of the comparisons in this paper (optical color, sSFR, B/T, petrosian radius, R50/R90, environmental measures) separately on the central and satellite galaxies.  When we compare BBRD centrals (or satellites) to mass-weighted samples of the parent centrals (or satellites) we find no qualitative changes with the results comparing the entire BBRD and parent samples.  For example, the p-value of $\Sigma_5$ of centrals (satellites) is [0.97, 0.77] ([0.78,0.12]) for the [hsSFR, lsSFR] parent samples (also split into centrals and satellites).

\subsection{Morphology}

\begin{figure}
\includegraphics[scale=0.53]{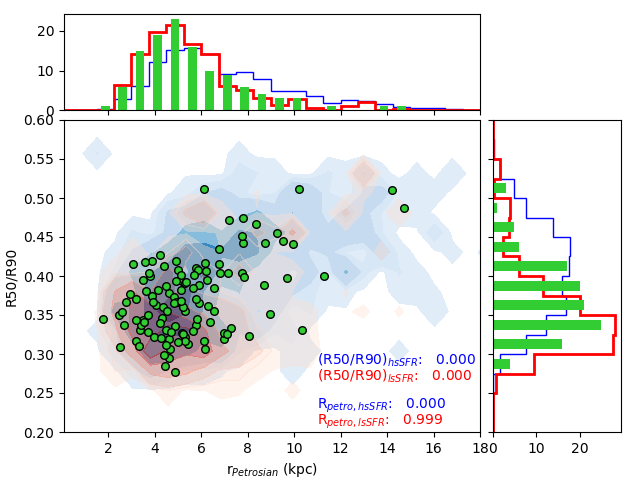}
\caption{The Petrosian radius (as defined by the SDSS project in \citet{2001AJ....121.2358B} and \citet{2001AJ....122.1104Y}.) is plotted versus the inverse galaxy concentration ratio (R50/R90), with colors and symbols as in Figure \ref{fig:parentsamplereddisk}. This ratio has shown to be a simple morphology discriminator, but our parent sample selection biases this tracer. Our sample of galaxies overlays the underlying distribution.\label{fig:petro}}
\end{figure}

\begin{figure}
\includegraphics[scale=0.53,trim=0mm 0mm 0mm 29mm,clip]{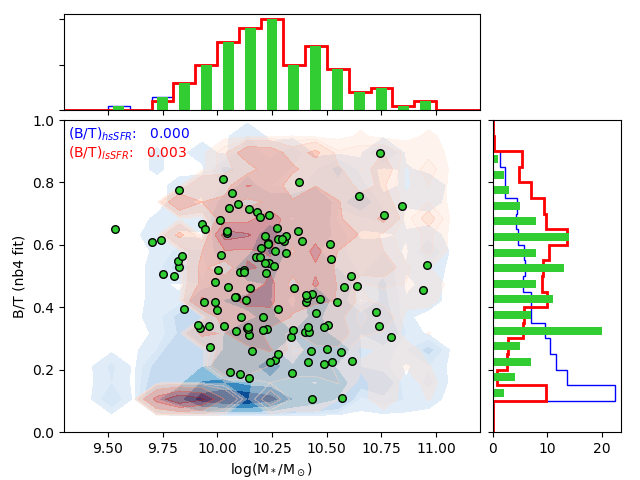}
\caption{The B/T ratio (using the nb4 bulge + disk fit) of the breakBRDs sample (green points) compared to the low- and high-sSFR parent samples (red and blue contours, respectively). Our galaxies are more bulge dominated than the high-sSFR parent sample.\label{fig:btcomp}}
\end{figure}

When we examine morphological measures of breakBRD galaxies in Figure \ref{fig:petro}, we see that these galaxies are mostly more concentrated (using the ratio of the radii with 50\% and 90\% of the Petrosian flux) and have small Petrosian radii, in better agreement with galaxies with low-sSFRs than those with high-sSFRs (again these values are from B04).  In fact, a KS test on the Petrosian radius distributions shows that the low-sSFR sample and breakBRD galaxies may come from the same distribution.  The evolution of these galaxies may be more complicated than their global (s)SFRs would indicate.
\begin{figure*}
\includegraphics[scale=1.,trim=0mm 0mm 0mm 20.0mm,clip]{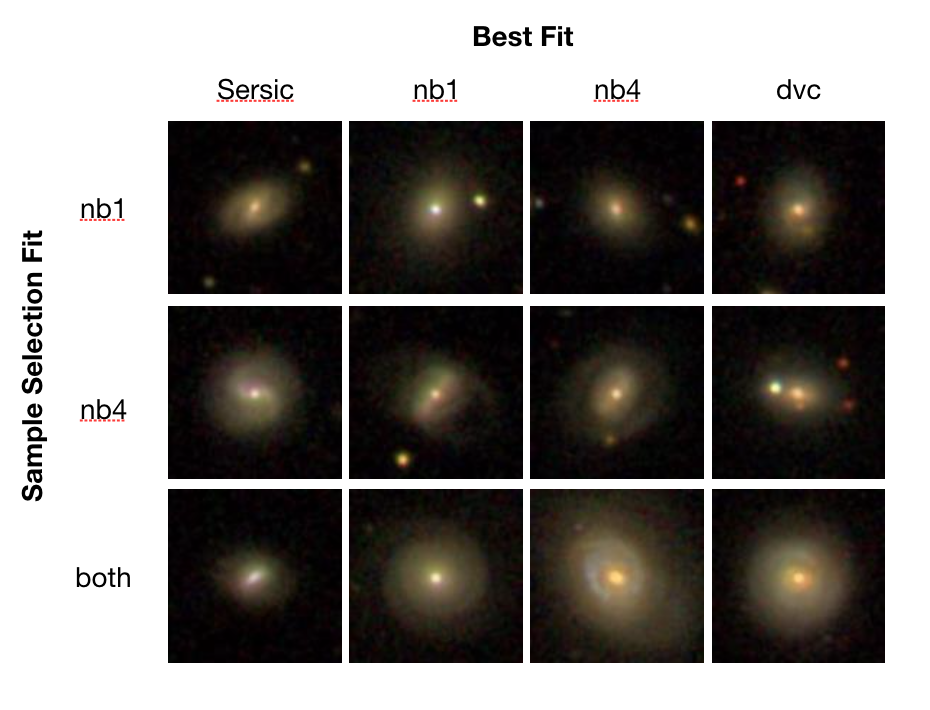}
\caption{Sample postage stamp images of breakBRD galaxies drawn from the SDSS imaging survey. The rows (labelled "Sample Selection Fit") indicate the fit which caused the galaxy to be selected for our sample (see Section \ref{sec:reddisks} and LG12 for more details). We note there are not morphological trends throughout the fits, which motivated our choice to include the broadest set to meet the criteria for sample membership. \label{fig:postagestamp}}
\end{figure*}

We next consider the morphologies of the breakBRD galaxies, and in particular whether their physical characteristics can give us insight into the causes of their central star formation.  We also consider whether they are morphologically more similar to the low- or high-sSFR parent samples.

We also show in Figure \ref{fig:btcomp} that if all galaxies are fit using the nb4 bulge + disk model, the bulges in our sample tend to be larger than most high-sSFR galaxies of similar masses. In fact, in our sample the lowest mass galaxies have the highest bulge-to-total ratio, while higher mass galaxies are more scattered.  The distribution of bulge-to-total ratios of breakBRD galaxies is significantly different from those of the parent samples (p-values $<<$ 0.01).

In fact, we note that if we only consider central galaxies, the p-values comparing breakBRD galaxies to the mass-weighted low-sSFR parent sample for R50/R90, petrosian radius, and B/T are all above 0.05.  Satellite galaxies only have large p-values for the petrosian radius.

We finally examine the images of all of the galaxies in our sample, first looking for bars.  Bars are frequently invoked when it comes to morphological transformation (e.g. \citet{2004ARA&A..42..603K}). In some circumstances they are found to induce central star formation (e.g. \citet{2017ApJ...848...87C, 2019MNRAS.484.5192C}) and it has long been thought that bars may encourage gas to flow centrally to enhance star formation in the bulge. However, this is an unsatisfactory explanation overall for our sample as the barred fraction (visually inspected by both authors) is roughly consistent with what has been measured for the spiral population (30\%-40\%, \citet{1993RPPh...56..173S, 2010ApJ...714L.260N}). It has also been argued that in galaxies with large bulges, bars may stabilize gas from collapse and suppress star formation \citep{2018MNRAS.474.3101J}.  

It is important to peruse the images to look for visual evidence of evolutionary mechanisms or a similarity in morphological type.  For example, in one of our 118 breakBRD galaxies we identify a late-stage merger.  As discussed above, we also search for bars. However, as shown in Figure \ref{fig:postagestamp}, this sample does not present as a single morphological type. All of the galaxies have a bright core, but some contain arms, some contain bars, and some have a smoother extended bulge appearance. In Figure \ref{fig:postagestamp} each row shows a subsample of galaxies that have been added to the breakBRD sample using the labeled fit.  In each row, however, we see barred galaxies, those with likely spiral structure, and those with smooth stellar distributions.  The bulge+disk fit that classified these galaxies as breakBRDs (the "sample selection fit") does not morphologically sort or differentiate the galaxies. 
Figure \ref{fig:postagestamp} highlights the diversity of breakBRD galaxies.

\section{Star Formation History} \label{sec:SFH}

\begin{figure}
\includegraphics[scale=0.53,trim=0mm 0mm 0mm 27.8mm,clip]{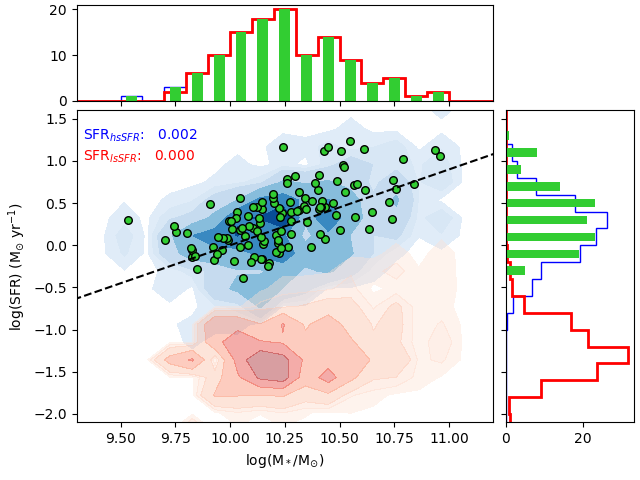}
\includegraphics[scale=0.53,trim=0mm 0mm 0mm 27.8mm,clip]{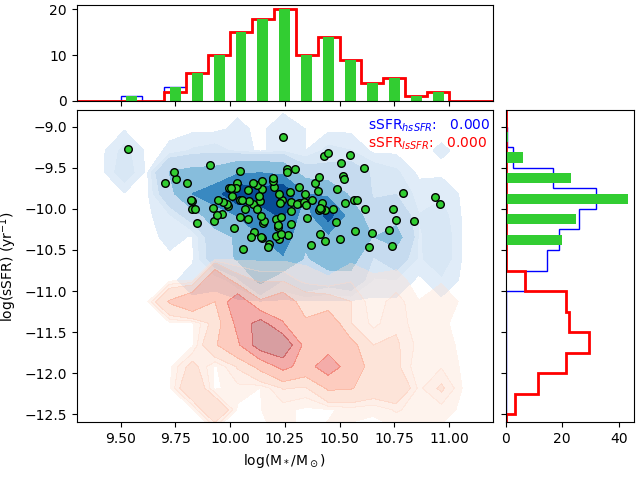}
\caption{The star formation rate (upper) and the specific star formation rate (lower) from B04 integrated across the entire galaxy. The colors and symbols are as in Figure \ref{fig:parentsamplereddisk}. In the upper panel we show the star-forming main sequence from Peng et al. (2010) as a black dashed line.  Our sample, as selected, is star-forming.\label{fig:massSFRs}}
\end{figure}

What can we discover about the star formation history of these galaxies from the multiwavelength data available? Using both sSFR and SFR, compared to the stellar mass, breakBRD galaxies are star-forming and situated in the blue cloud, as is expected for spiral star-forming galaxies. We begin by identifying the location of these galaxies compare to the galaxy star-forming ``main sequence”, with Figure \ref{fig:massSFRs} showing they fall firmly along this sequence. We use the (s)SFRs from B04, updated to better perform aperture corrections and account for dust attenuation\footnote{https://wwwmpa.mpa-garching.mpg.de/SDSS/DR7/sfrs.html}.  However, we note the the KS test indicates that the sSFRs and SFRs of the breakBRD galaxies are not drawn from the same distribution as the star-forming parent sample.

We now examine the star formation history of the sample by examining the ultraviolet, optical, and infrared color-color space to identify trends in star formation from the recent past (100 Myrs, probed by the ultraviolet) to the more distant (Gigayear timescales, measured by the infrared). We use archival data from GALEX (The Galaxy Evolution Explorer, \citep{2003SPIE.4854..336M}), WISE (the Wide-field Infrared Survey Explorer, \citep{2010AJ....140.1868W}), and SDSS, to explore these relationships. We also find a small subsample of these galaxies ($\sim10$) that have been observed in the ALFALFA HI survey, and use them to investigate possible relations with the neutral gas properties.

\subsection{Galaxy Scale Star Formation History}
We progress through the star formation history from most recent star formation as represented by the ultraviolet to the oldest as measured in the WISE infrared bands. We in particular look for signs of a quenching population or star formation which is being slowed in recent times, as we investigate the the color-color relationships throughout wavelength space.

\subsubsection{GALEX/UV}

We matched our sample with the GALEX archive, via the NASA-Sloan Atlas (NSA)\footnote{http://www.nsatlas.org/} to ensure consistent photometry.  The NSA combines reprocessed SDSS DR8 results using improved background subtraction \citep{2011AJ....142...31B} with GALEX Near and Far UV observations self consistently. We use the UV as an indicator of recent star formation, in the last 100 Myr, although some recent work has shown there may be contamination of the star formation signal by blue horizontal branch stars \citep{2018MNRAS.476.1010A}.

\begin{figure}
\includegraphics[scale=0.53]{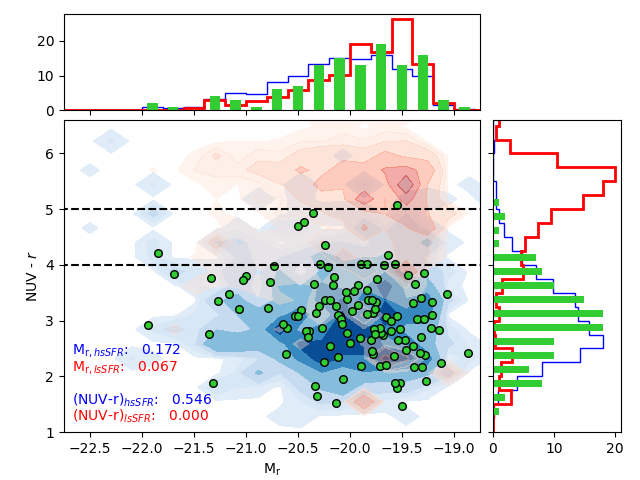}
\caption{NUV - r versus M$_r$ of our sample over plotted on the comparison sample, with the underlying contours representing the high- and low-sSFR subsamples (in blue and red, respectively). Black dashed lines denote the ``green valley" \citep{2014SerAJ.189....1S}.  The NUV-r color distribution more strongly agrees with that of the high-sSFR parent sample.  Both NUV and r magnitude values shown are drawn from the NASA-Sloan Atlas$^1$. \label{fig:nuv2}}
\end{figure}

 The results are shown in Figure \ref{fig:nuv2}. BreakBRD galaxies largely fall bluer than the ``green valley", defined here as $4 < NUV-r < 5$ \citep{2014SerAJ.189....1S}.  This more tightly constrains the star formation to the more recent history than the D$_{n}4000$ fiber measurements used to select our sample. 
 
 We see in the parent sample that the low- and high-sSFR samples are fairly well-separated in NUV-r color space.  In agreement with Figure \ref{fig:massSFRs}, breakBRD galaxies all have high-sSFRs, and the distribution of NUV-r colors is similar to that of the high-sSFR sample.  
 
\begin{figure}
\includegraphics[scale=0.53,trim=0mm 0mm 0mm 29.5mm,clip]{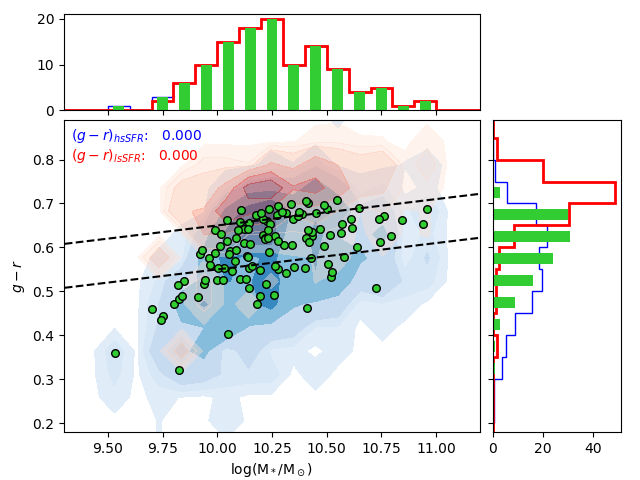}
\includegraphics[scale=0.53,trim=0mm 0mm 0mm 29.5mm,clip]{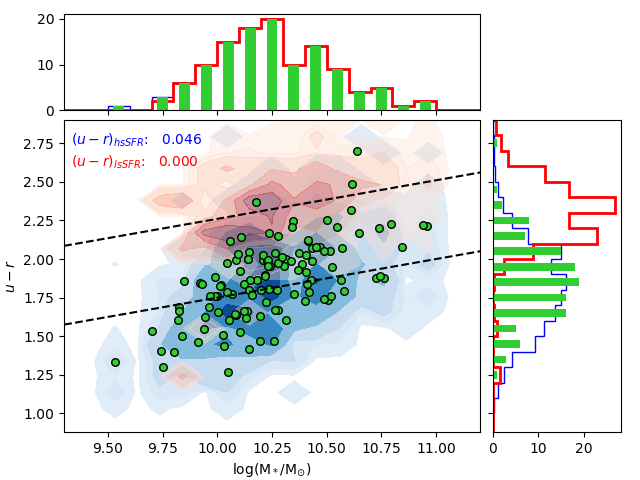}
\caption{Color-stellar mass diagrams comparing the breakBRD sample to the low- and high-sSFR parent sample (colors and symbols as in Figure \ref{fig:parentsamplereddisk}). The top and bottom panels are the $g - r$ and $u - r$ colors, respectively. The black dashed lines denote the ``green valley" transitional region \citep{2013MNRAS.429.2212M,2014MNRAS.440..889S}.  Although our galaxies are chosen to have red disks ($g -r$ $>$ 0.655), the total galaxy colors are bluer, indicating star formation in the central regions.  In both panels, our sample tends to have transitional or blue colors. \label{fig:ccgr}}
\end{figure}

\subsubsection{Optical Colors}

Using global optical $g - r$ and $u -r$ colors from B04 (Figure \ref{fig:ccgr}), we find that breakBRD galaxies primarily reside in the green valley \citep{2013MNRAS.429.2212M,2014MNRAS.440..889S}, with a strong tail in the blue star-forming region.  The distribution of these galaxies in optical color space indicates that this population of galaxies may be transitioning from blue to red.

 In detail, the $g - r$ colors of a significant minority of galaxies fall red of the transition region, while only 3 galaxies are redder than the transition region in $u - r$ colors. However, if we consider the distribution of the parent sample, we see that even some galaxies with high-sSFRs (blue) have red $g - r$ colors, while conversely, galaxies with low-sSFRs (red) have transitional $u - r$ colors.  The distribution of optical colors of our sample lies between that of the high- and low-sSFR parent populations, supporting our hypothesis that these are a transitional population. 
  
We point out that in order to select our sample, we require the disk component of the LG12 galaxies to have $g -r$ $>$ 0.655.  As is clear from the upper panel of Figure \ref{fig:ccgr}, most colors from B04 are bluer.  This indicates that the stellar populations in the central region are likely bluer (younger) than in the disk.  In a single stellar population model of \citet{2005MNRAS.362..799M}, $g -r$ $>$ 0.655 implies stellar ages of $\geq$1 Gyr, but the ages of stars in the central regions are likely to be significantly lower.

In future work we will use spatially resolved star formation indicators to determine the local star formation history of these galaxies.

\subsubsection{WISE Colors}\label{sec:WISE}

\begin{figure}
\includegraphics[scale=0.53,trim=0mm 0mm 0mm 30mm,clip]{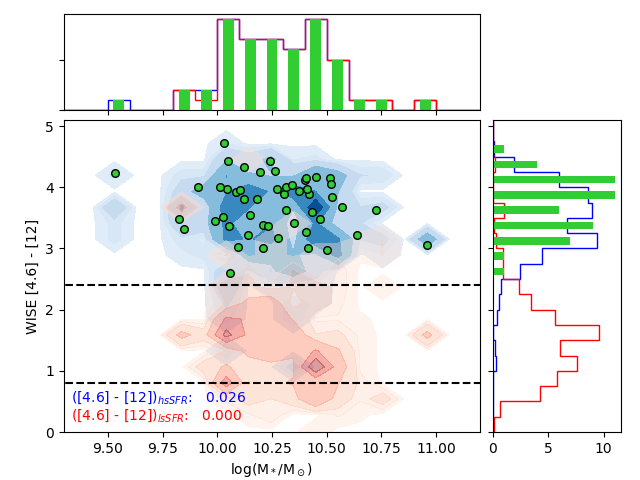}
\caption{WISE [4.6] - [12] color versus galaxy stellar mass (colors and symbols as in Figure \ref{fig:parentsamplereddisk}). The InfraRed Transition Zone is labeled with black dashed lines.  BreakBRD galaxies all lie in the star-forming region of this diagram.   \label{fig:irmstar}}
\end{figure}

We next examine our sample in WISE color (drawn from the VAGC \citep{2005AJ....129.2562B, 2009ApJS..182..543A}), to see where the galaxies lie with respect to the InfraRed Transition Zone (IRTZ, \citet{2014ApJ...794L..13A}).  The IRTZ is a proposed region of infrared color space designating a split between early and late type galaxies. It is proposed to contain galaxies primarily finishing their move through the optical green valley. We match the positions of the breakBRD galaxies and those from the parent sample with WISE sources using the astropy skycoords package \citep{2013A&A...558A..33A, 2018AJ....156..123A}. 51 of our galaxy sample and 1100 galaxies from the parent sample have WISE magnitudes greater than zero (552/548 with high/low-sSFRs).  

In Figure \ref{fig:irmstar} we plot the WISE [4.6] - [12] color versus galaxy stellar mass.  All of our sample galaxies lie in the star-forming region of this diagram, and have not begun transitioning through IR color space. As discussed in \citet{2014ApJ...794L..13A}, blue [4.6] - [12] color indicates that the galaxies retain ISM. The physical distribution of the ISM is, however, unknown. 

\begin{figure}
\includegraphics[scale=0.53]{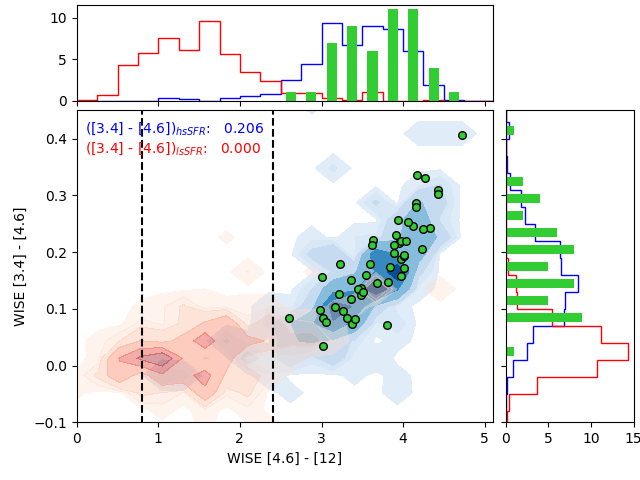}
\caption{The WISE [3.4] - [4.6] versus [4.6] - [12] diagram.  The colors are consistent with those of star-forming spirals (Section \ref{sec:WISE}).}  
\label{fig:ircolcol}
\end{figure}

Figure \ref{fig:ircolcol} is the WISE [3.4] - [4.6] versus [4.6] - [12] diagram.  Again, we see that these galaxies mostly fall in the star-forming region.  As with WISE [4.6] - [12] colors, a KS test finds that the WISE [3.4] - [4.6] colors of the breakBRD galaxies may be drawn from the same distribution as those of the high-sSFR parent sample
. Indeed, according to \citet{2010AJ....140.1868W} and \citet{2011ApJ...735..112J}, the breakBRD galaxies' [3.4] - [4.6] colors are bluer than AGN, and the [4.6] - [12] colors indicate that these are star-forming spiral galaxies \citep{2017ApJ...836..182J}.  In fact, some of the galaxies have [4.6] - [12] colors indicating that they may be starburst galaxies.

\begin{figure}
\includegraphics[scale=0.53]{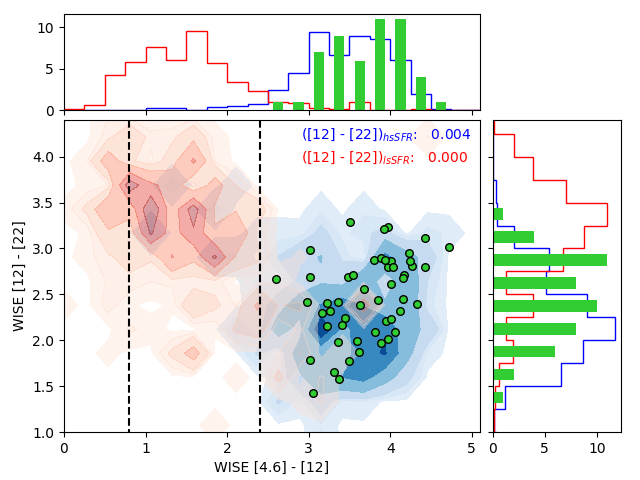}
\caption{The WISE [12] - [22] versus [4.6] - [12] color-color diagram. The [12] - [22] colors indicate that these galaxies are unlikely to be ellipticals. \label{fig:ircolcol2}}
\end{figure}

Figure \ref{fig:ircolcol2} is the WISE [12] - [22] versus [4.6] - [12] diagram.  The breakBRD galaxies reside in the star-forming region in this diagram as well.  The [12] - [22] colors tend to be bluer than 3.0, indicating that these galaxies are unlikely to be ellipticals (luminous red galaxies; \citet{2014MNRAS.442.3361N}).

\begin{figure}
\includegraphics[scale=0.53]{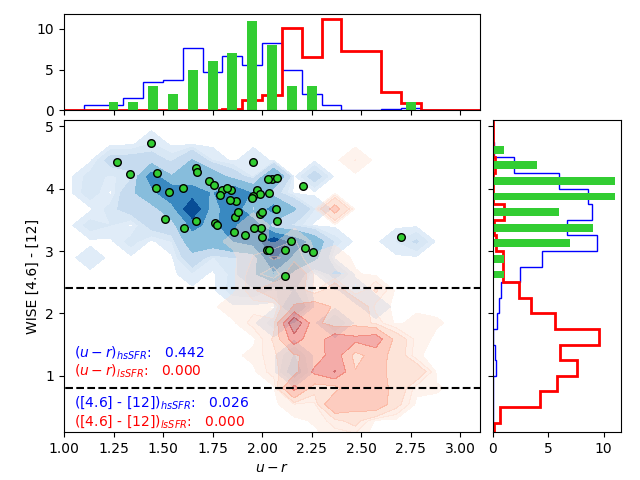}
\caption{WISE [4.6] - [12] versus $u - r$ color.  These galaxies have blue WISE colors, and lie somewhat above the distribution of the Galaxy Zoo sample (see \citet{2014ApJ...794L..13A}). }\label{fig:WISE_umr}
 
\end{figure}

Finally, in Figure \ref{fig:WISE_umr} we directly compare the WISE [4.6] - [12] and $u - r$ colors.  The galaxies in our sample range from star-forming to transitioning in the green valley in $u - r$ (see Figure \ref{fig:ccgr}), but are all above the IRTZ in [4.6] - [12].  If we follow the logic of \citet{2014ApJ...794L..13A}, that galaxies first transition in optical colors and then WISE colors (their Figure 1), we may be identifying galaxies very early in their transition from star-forming to quenched.  While their SFRs have decreased, they still retain some ISM, allowing for $[$12 $\mu$m$]$ emission.

\subsubsection{ALFALFA/HI}\label{sec:HI}

\begin{figure}
\includegraphics[scale=0.56]{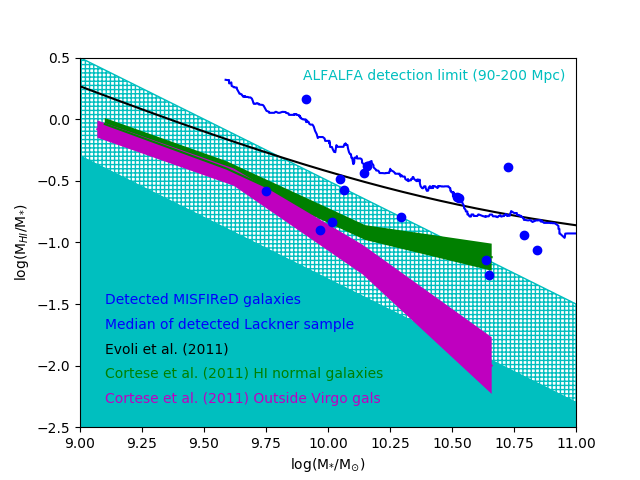}
\caption{The HI mass to stellar mass ratio as a function of stellar mass for the 16 galaxies with HI detections (blue points).  The blue line is the running mean of the M$_{HI}$/M$_*$ of the parent sample that has HI detections.  The black line is the M$_{HI}$/M$_*$ fit from \citet{2011ApJ...743...45E}.  The green and magenta regions are the M$_{HI}$/M$_*$ fractions in four mass bins in \citet{2011MNRAS.415.1797C}.  The cyan regions estimate the ALFALFA detection limit starting at a distance of 90 Mpc (hatched region) to 200 Mpc (solid region). \label{fig:MHI_Mstar}}
\end{figure}

Given that the WISE colors indicate that our galaxies may still have significant mass in the ISM, we compare our sample to the ALFALFA data release \citep{2018ApJ...861...49H,2011AJ....142..170H,2005AJ....130.2598G}.  Although we have a number of galaxies outside the ALFALFA survey region, about 67 of our galaxies are within the ALFALFA footprint.  Of those galaxies, 16 have HI detections, as found by matching optical counterparts identified in the ALFALFA catalog.  We also visually inspected cutouts of the sky around these galaxies, and found that all galaxies with counterparts within 0.001 degrees of our sample were a match.  This is similar to the fraction of galaxies in the parent sample that have HI detections.  

In Figure \ref{fig:MHI_Mstar} we plot M$_{HI}$/M$_*$ as a function of M$_*$ for the sample galaxies with HI detections as the blue points.  We also plot the running mean (median gives nearly identical results) of the M$_{HI}$/M$_*$ for the parent sample galaxies with HI detections as the blue line.  

The \citet{2011MNRAS.415.1797C} sample is from the Herschel Reference Survey, which consists of 322 galaxies \citep{2010PASP..122..261B}.  \citet{2011MNRAS.415.1797C} found that 305 of the galaxies had been observed in HI, with 265 detections.  \citet{2011MNRAS.415.1797C} created three samples:  two defined by the environment (inside or outside of the Virgo cluster) and one that included all HI normal galaxies (those with at least 30\% of the HI of isolated galaxies with the same diameter and morphological type).  Here we show the mean trends including error bars in four mass bins for the HI normal (green) and outside Virgo (red) samples.  

\citet{2011ApJ...743...45E} use a different method to determine M$_{HI}$/M$_*$ as a function of stellar mass.  They use the mass-ranking method introduced in \citet{2004MNRAS.353..189V}, assuming that the mass of HI is an increasing monotonic function of the mass of the stellar disk.  They can then use the galaxy stellar mass function from \citet{2010MNRAS.404.2087B}, and the HI mass function from \citet{2005MNRAS.359L..30Z} to determine the HI-to-stellar mass ratio.  We plot this fit with the black line in Figure \ref{fig:MHI_Mstar}.  

Finally, the cyan filled and hatched regions roughly denote the ALFALFA detection limit for galaxy at a distances of 200 or 90 Mpc, respectively, based on the Spaenhauer diagram we made from the the full ALFALFA data release ($\alpha$.100 catalog).

We first highlight that breakBRD galaxies are not universally gas-rich.  Although the WISE colors of the majority of these galaxies indicate that they have star-forming gas, only about 25\% of the galaxies in the ALFALFA footprint have HI detections.  While the ALFALFA detection limit may play a role in the lack of detections, this indicates that the HI reservoir is low for our sample of galaxies.  These galaxies tend to be more gas-poor than the parent sample, although we have not corrected for the distance distribution of galaxies.  Our sample M$_{HI}$/M$_*$ is scattered around the \citet{2011ApJ...743...45E} relation.  Although ALFALFA detected galaxies in our sample lean towards being more gas-rich than the \citet{2011MNRAS.415.1797C} samples, we note that their samples were much closer, so they could detect much lower HI masses and they included non-detections in their gas richness calculations.  

The sample is not a gas-rich population.  We need deeper observations to determine the gas reservoirs available to these galaxies.

\subsection{Central Star Formation History}\label{sec:fiber}

As discussed in Section \ref{sec:sample}, our galaxies were chosen to have D$_n$4000 measures in the central fiber indicating recent star formation, within the last 1 Gyr.  Here we more carefully consider the central star formation in these galaxies.

\begin{figure}
\includegraphics[scale=0.53]{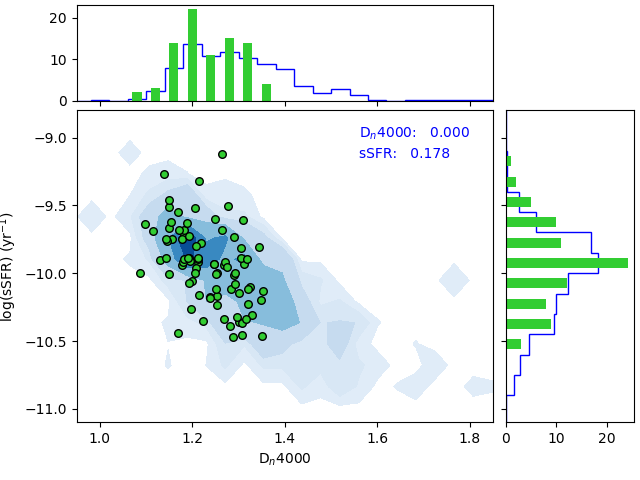}
\caption{The D$_n$4000 measure in the central fiber compared to the total sSFR from B04 for the star-forming subsamples of both the breakBRD and high-sSFR parent samples.  Our galaxies are similar to the parent star-forming sample, but lack the high D$_n$4000 values in the fiber. \label{fig:d4000sSFRtot}}
\end{figure}

We use the values for D$_n$4000 and the fiber and total (s)SFRs from B04.  As discussed in that paper, the emission lines can only reliably be used to determine the SFR of galaxies that lie in the star-forming region in the BPT diagram.  Therefore, in this section we only consider those galaxies.  This leaves us with a sample of 85 breakBRD galaxies and 944 comparison galaxies from the parent sample (galaxies selected using both the BPT diagram and with a total sSFR $>$ 10$^{10.9}$).  We will call these the star-forming samples.  

In Figure \ref{fig:d4000sSFRtot}, we compare the D$_n$4000 - total sSFR relation for the star-forming samples of breakBRD and parent sample galaxies.  We see that the central D$_n$4000 measure is similarly correlated with the galaxy-wide sSFR for both samples.  However, there is a larger tail of high central D$_n$4000 in the comparison star-forming sample that is not found in our sample.

\begin{figure}
\includegraphics[scale=0.53]{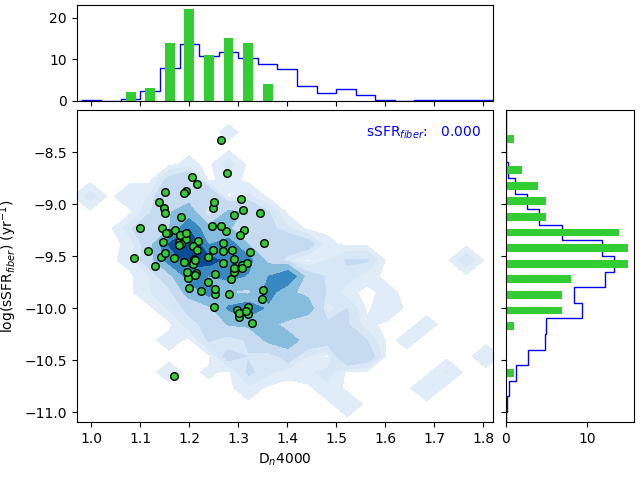}
\caption{The D$_n$4000 measure in the central fiber compared to the fiber sSFR from B04 for the star-forming subsamples of the breakBRD and high-sSFR parent samples.  Our galaxies mostly have higher central sSFRs than the parent star-forming sample. \label{fig:d4000sSFRfib}}
\end{figure}

When we focus on the sSFR within the fiber in Figure \ref{fig:d4000sSFRfib}, we begin to see differences between the star-forming galaxies in the breakBRD and parent samples.  We highlight that the fiber sSFR for star-forming galaxies is based on the emission lines, mostly H$\alpha$ (B04).  This is therefore a good comparison of the recent star formation in the center of these galaxies. The sSFR within the fiber of our galaxies tends to be higher than in the parent sample.  Together with Figure \ref{fig:d4000sSFRtot}, this indicates that for a particular D$_n$4000, more of the current star formation is in the central region of our galaxies.  

\begin{figure}
\includegraphics[scale=0.53,trim=0mm 0mm 0mm 30mm,clip]{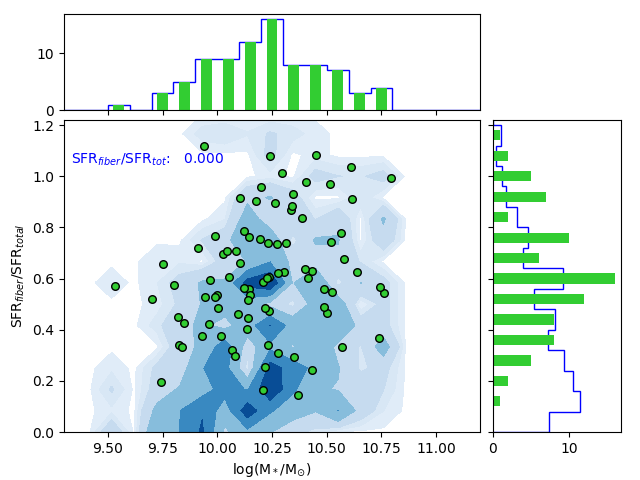}
\caption{The ratio of the fiber SFR to the total SFR as a function of stellar mass for the star-forming subsamples of the breakBRD and high-sSFR parent samples.  As a population, the breakBRD galaxies have much more centrally-concentrated star formation.} \label{fig:SFRfib_vs_tot}
\end{figure}

Finally, the central concentration of star formation is explicitly shown in Figure \ref{fig:SFRfib_vs_tot}.  Note that the values we use are the mean from the likelihood distribution of SFRs (B04), which means that in a few cases the mean fiber SFR could be higher than the mean total SFR (using the median values does not change the distributions).  breakBRD galaxies have a dramatically different distribution, with a higher fraction of their star formation in the central fiber compared to the total star formation.  This clearly shows a significant difference between star-forming breakBRDs and star forming galaxies in the parent sample. We note that if we select star-forming galaxies with red disks and high fiber sSFRs ($>$10$^{-10}$ yr$^{-1}$), only the distributions of D$_n$4000 and the disk colors differ significantly from the star-forming breakBRD sample.  This gives even more support to using D$_n$4000 as a tracer of ``recent" star formation.

Comparing the D$_n$4000 to SFRs from emission lines can be used to indicate whether these galaxies have been quenched within the last 100 Myr to Gyr.  Emission lines show quite recent star formation, within tens of Myr, rather than the larger 1 Gyr window indicated by the D$_n$4000, so recently quenched galaxies may have low D$_n$4000 values and low-sSFRs.  By this measure, our galaxies have likely not been quenched in their central regions.

\section{Discussion}\label{sec:disc}

Our sample was chosen to have red disks and central star formation.  When we examine the global colors of these galaxies, they present results that run somewhat counter to our expectations for their morphologies and environments.  In particular, only in optical colors do these galaxies largely fall into a transitional color region.  In both UV - optical and WISE colors these galaxies appear to be in the star formation region.

This may appear puzzling, given that optical colors trace older star formation than UV - optical. However, as discussed in \citet{2014SerAJ.189....1S}, UV - optical is sensitive to lower sSFRs, so low current star formation could be driving the blue UV colors. The blue WISE colors are less of a mystery, as \citet{2014ApJ...794L..13A} show that galaxies will transition in optical colors before WISE colors, and the blue WISE colors indicate the presence of ISM gas. We stress that these data do not demonstrate whether these galaxies are transitioning from star-forming to quenched:  in many parameters discussed in this paper, breakBRDs fall between high- and low-sSFR galaxies. 

As we have discussed, the classic picture of merger-driven galaxy growth is inside-out: older bulges surrounded by star-forming disks. 
In late-type galaxies, when we observe cessation of star-formation in the disk we might expect environmental effects to be at play. However, not only do breakBRDs have continued star formation in their centers, but their central sSFRs are higher than those of star-forming galaxies from the parent sample.  There are several known scenarios in which central star formation is expected, and we discuss them now.  

First, bars may induce central star formation \citep{1993A&A...268...65F, 1997A&A...323..363M, 2004ARA&A..42..603K, 2017ApJ...848...87C}.  However, our sample has 30-40\% bar fraction (categorizing optical images visually), which seems inadequate to explain the observations. However, it is possible that bars have been weakened or destroyed through processes that drive gas towards the center. This loss of angular momentum and inflow will create a high central mass concentration (CMC) and destroy the bar \citep{1993A&A...268...65F, 2004ARA&A..42..603K}.  In the future we will examine the velocity dispersion in the breakBRD bulges to search for the imprint of a CMC \citep{2005MNRAS.363..496A}. 

Observations have found that low-mass galaxies may have central star formation \citep{2008ApJ...682L..89G, 2013MNRAS.436..259P}, and this has been well-documented in simulations \citep{2010Natur.463..203G, 2016ApJ...820..131E}. BreakBRD galaxies appear to fall outside this paradigm and have a wide range of stellar masses, with most of the population having stellar masses above 10$^{10}$ M$_\odot$.  We note that \citet{2016ApJ...820..131E} predict that episodes of strong central star formation will correspond with small effective radii.  While in this work we examine the Petrosian radius, we find no clear correlation between sSFR and galaxy radius (even when we only consider the $~$20 galaxies with stellar masses less than 10$^{10}$ M$_\odot$).  Carefully studying nearby galaxies with higher masses that exhibit this centralized mode of star formation, i.e. breakBRDs, will allow us to examine the factors that specify the locations of star formation in galaxies. 

Focusing on observations of more massive galaxies, many recent studies of S0s find that their last star formation episode was in the bulge, whether they reside in the field or clusters \citep{2001ApJ...563..118P, 2006A&AT...25..199S, 2012MNRAS.427..790S, 2012MNRAS.422.2590J, 2013MNRAS.428.1296J}. However, this may not be universal, as \citet{2015AJ....150...24K} find that in isolated lenticular galaxies the ages of the bulge and disk are very similar. \citet{2018MNRAS.474.1909F} find that for galaxies more massive than 10$^{10}$ M$_{\odot}$, bulges tend to be older than disks, while at lower masses the opposite is true.  Most of these studies, that specifically selected S0s, find that both the bulge and disk ages are more than 1 Gyr, in contrast to the young ages indicated by the emission-line fiber SFRs. In the future we will test whether breakBRD galaxies are likely to be in their final star formation episode by searching for their remaining reservoir of gas.  As we have shown in Section \ref{sec:HI}, ALFALFA data does not give us enough insight into whether our sample retains a significant reservoir of HI gas for future star formation.

One expectation for breakBRD galaxies would be that spiral galaxies with red disks were likely those that had experienced ram pressure stripping or another environmental effect that removed or disrupted the galaxy's gas reservoir. The lack of evidence for an environmental trend suggests either that this is not a primary effect or that several different effects result in similar morphological outcomes.

BreakBRD galaxies do not clearly fall into any well-studied category for centrally-concentrated star formation.  They are not all satellites, barred, low-mass, or clearly quenching.  Luckily, our sample is small enough that we can study each galaxy in detail instead of requiring an overarching explanation. This work is being followed up with optical integral field spectroscopy to better understand the spatial distribution of any recent star formation and how that correlates with the global properties discussed here. The diversity of this sample raises the question of whether central growth within red disks is unusual or if it is a short-lived stage in the life of most galaxies.  We are currently examining breakBRD analogs in IllustrisTNG to address this question (Kapferer et al., in prep).

\section{Conclusions}\label{sec:conclusion}

In this paper we have introduced breakBRD galaxies, a newly identified sample of galaxies that consists of local face on galaxies demonstrating central star formation using the D$_n$4000 within red ($g - r > 0.655$) disks.  We use the BPT diagram to select only star-forming and composite galaxies.  We have shown:

1) breakBRDs are distributed across a large range of stellar mass, indicating that mass-dependent processes that drive central star formation are not universally active (Figure \ref{fig:massden}).

2) Our sample galaxies are well-distributed across environments in a similar fashion to the parent sample (Figures \ref{fig:massden} \& \ref{fig:rho5cd}).  This implies either a process that is not moderated by environmental factors, or several processes resulting in centralized star formation.

3) The NUV - r colors indicate that these galaxies have enough star formation to be in the star-forming sequence and not transitioning in this color space (Figure \ref{fig:nuv2}).

4) These galaxies reside in the optical green valley, with a significant tail in the blue star-forming region of the color-magnitude diagrams (Figure \ref{fig:ccgr}). 

5) Our sample galaxies have WISE IR colors that lie firmly in the star-forming galaxy region, with no galaxies in the IRTZ (Figures \ref{fig:ircolcol} \& \ref{fig:ircolcol2}).  This may indicate that there is still ISM in these galaxies. 

6) The ALFALFA data shows that some (16/67) of these galaxies have a gas reservoir available for future star formation (Figure \ref{fig:MHI_Mstar}).

7) Our selection of galaxies using D$_n$4000 and $g - r$ broadband colors has found galaxies that are currently forming stars within the central fiber (Figure \ref{fig:d4000sSFRtot}). The star formation, measured using the emission lines, is more centrally concentrated in these galaxies than in the parent sample (Figures \ref{fig:d4000sSFRfib} \& \ref{fig:SFRfib_vs_tot}).

\vspace{0.5cm}

We have separated our parent sample into galaxies with high- and low-sSFR.  Although the current sSFR of breakBRD galaxies are within the high-sSFR peak distribution (Figure \ref{fig:massSFRs}), we cannot currently determine whether these galaxies are quenching with their final star formation in the center, or whether we have selected galaxies that are currently preferentially growing their centers in stochastically-distributed star formation.  As we discuss in Section \ref{sec:disc}, either of these possibilities is of great interest for understanding what determines the distribution of star formation in galaxies. Future papers will explore the nature of the spatially resolved star formation as well as the star formation histories of the sample.

In summary, breakBRD galaxies do not seem to be quenching satellites.  Indeed, this sample does not lend itself to a single unifying explanation.  As shown in Figure \ref{fig:postagestamp}, our galaxies span a range of morphologies.  Because our sample is $\sim$100 galaxies, in future work we will examine each galaxy in detail to determine what properties other than their star formation distribution are unique and may play a role in causing central star formation in red disk galaxies.

\bibliography{bluebrd}{}

\acknowledgments
The authors would like to acknowledge Jenny Greene, Julianne Dalcanton, and Jessica Werk who all provided valuable feedback for the manuscript. They would also like to thank Dan Foreman-Mackey and Mehmet Alpaslan for discussions about mass-weighting samples.

This material is based upon work supported by the National Science Foundation under Grant No. 1813462.

Funding for the SDSS and SDSS-II has been provided by the Alfred P. Sloan Foundation, the Participating Institutions, the National Science Foundation, the U.S. Department of Energy, the National Aeronautics and Space Administration, the Japanese Monbukagakusho, the Max Planck Society, and the Higher Education Funding Council for England. The SDSS Web Site is http://www.sdss.org/.

The SDSS is managed by the Astrophysical Research Consortium for the Participating Institutions. The Participating Institutions are the American Museum of Natural History, Astrophysical Institute Potsdam, University of Basel, University of Cambridge, Case Western Reserve University, University of Chicago, Drexel University, Fermilab, the Institute for Advanced Study, the Japan Participation Group, Johns Hopkins University, the Joint Institute for Nuclear Astrophysics, the Kavli Institute for Particle Astrophysics and Cosmology, the Korean Scientist Group, the Chinese Academy of Sciences (LAMOST), Los Alamos National Laboratory, the Max-Planck-Institute for Astronomy (MPIA), the Max-Planck-Institute for Astrophysics (MPA), New Mexico State University, Ohio State University, University of Pittsburgh, University of Portsmouth, Princeton University, the United States Naval Observatory, and the University of Washington.

\vspace{5mm}
\facilities{Arecibo, GALEX, Sloan, WISE}
\software{Astropy,\footnote{http://www.astropy.org} a community-developed core Python package for Astronomy.}

\end{document}